\documentclass[11pt,a4paper]{article}
\usepackage[T1]{fontenc}
\usepackage[utf8]{inputenc}
\usepackage{lmodern}
\usepackage[margin=27mm]{geometry}
\usepackage{amsmath,amssymb,amsthm,booktabs,array,tabularx,microtype,hyperref,xcolor}
\hypersetup{colorlinks=true,linkcolor=blue!55!black,urlcolor=blue!55!black,citecolor=blue!55!black}
\newtheorem{proposition}{Proposition}
\newtheorem{theorem}{Theorem}
\newtheorem{definition}{Definition}
\newtheorem{remark}{Remark}

\newcommand{\unknown}{\bot}

\title{Adaptive Allele--Haplotype Incidence Indexing for Pangenome Queries:\\Representation, Reconstruction, and Benchmarking}
\author{Roberto Garrone\\Department of Informatics, Systems and Communication\\University of Milano-Bicocca, Italy}
\date{October 28, 2026}
\usepackage{graphicx}
\begin{document}
\maketitle
\begin{abstract}
Population-scale analysis of phased genetic variants requires both allele-carrier retrieval and ordered haplotype context. We examine H1, an adaptive allele--haplotype incidence-row index, and H2, a canonical edge-path representation derived from a fixed, nonoverlapping variant catalog. We prove conditional reconstruction of abstract H2 walks and distinguish this property from unrestricted equivalence to genomic variation graphs. A completed local notebook experiment used a chromosome 1 interval (20--22 Mb), 200 diploid individuals, 24,921 SNV/indel records, and 45 SVTYPE-annotated records. For 24,966 extracted nonreference rows across 400 haplotypes, the serialized H1 incidence-row files occupied 0.872 MiB with a simple adaptive layout and 1.126 MiB with a custom Roaring-like layout, versus 1.190 MiB for a bitmap-payload baseline; a whole-file Zstandard archive of the adaptive layout occupied 0.207 MiB. For a separately constructed nonoverlapping SNV projection containing 22,727 sites, indexed H2 edge sequences occupied 75,130,882 bytes in a raw format and 5,156,418 bytes with context-Huffman encoding (14.57-fold reduction). Carrier-equivalence and graph-adjacency assertions passed in the notebook, but full cross-callset phasing, nucleotide-level SV reconstruction, and comparison with production indexes remain unresolved. The measurements establish a restricted implementation feasibility result, not superior end-to-end pangenome compression.
\end{abstract}

\section{Introduction}
Representations for large sets of related genomes must negotiate three requirements: storing allelic variation, preserving enough structural context to interpret genomic alternatives, and answering queries without repeatedly scanning individual genotypes. The variant call format (VCF) and its binary counterpart (BCF) provide a practical site-oriented interchange standard; other systems index genotype sets or graph-embedded haplotypes directly \cite{danecek2011,gqt2016,bgt2016,gbwt2020}. A human pangenome reference makes diverse genomic sequences available as paths in a graph, but the choice of a population-wide representation affects both memory and query performance \cite{eizenga2020,liao2023,baaijens2022}.

It would be incorrect to describe an allele-by-haplotype binary matrix as intrinsically new. Compressed genotype tables, succinct bitmaps, haplotype indexes, and recent graph-based genotypic representations already use related information and exploit population regularities \cite{gqt2016,bgt2016,grg2025,kodama2026}. The methodological tension motivating the present work is narrower: a scheme that is simple for allele carrier enumeration may be inefficient for ordered haplotype queries, while path-optimized structures may entail different construction and indexing costs. Whether an adaptive row layout offers a useful trade-off is an empirical question.

We call the per-allele membership view H1 and a reference-coordinate, edge-walk view H2. Their relationship depends on the accompanying variant catalog and phasing information. We therefore ask: \emph{under what restrictions can a fixed catalog and exact incidence map be transformed into contiguous haplotype walks, and when do the associated storage and query costs justify maintaining an allele-oriented index?}

The present study makes three bounded contributions. First, it specifies the H1 incidence data model and distinguishes idealized encoding costs from actual serialized row stores. Second, it establishes conditional reconstruction of H2 under a reference-anchored, nonoverlapping-interval model and characterizes failure conditions. Third, it reports a local, reproducible-in-principle implementation pilot, including physical file-size measurements, path-consistency checks, and limitations of the candidate baselines. External production-index comparisons and whole-genome validation are not claimed.

\section{Related work and differentiation}
\subsection{Genotype and haplotype indexes}
GQT and BGT demonstrate that compact genotype stores can support direct queries over large cohorts \cite{gqt2016,bgt2016}. The positional Burrows--Wheeler transform exploits haplotype similarity for matching and storage \cite{durbin2014}; the graph BWT extends efficient indexing to paths embedded in graphs \cite{gbwt2020}. These are substantive baselines, not merely alternate file formats. For H1, carrier enumeration by row inspection and frequency estimation are ordinary index operations, not by themselves new methods.

\subsection{Variation graphs and contemporary compression}
Variation graphs are not uniquely determined by a genotype incidence relation. Their vertices, labels, oriented edges, distinguished walks, and construction objectives can differ while explaining the same variants \cite{baaijens2022,eizenga2020}. Modern genotype representation graphs (GRGs) losslessly encode phased whole-genome polymorphisms and enable computation without materializing a conventional matrix \cite{grg2025}. A 2026 preprint describes a further GRG implementation \cite{grgv22026}. A separate 2026 preprint presents \emph{kodama}, which exploits genealogical relatedness to compress genotypes and accelerate linear algebra \cite{kodama2026}. Accordingly, superiority over an uncompressed bitmap is insufficient evidence of a new state of the art.

H1 differs operationally when it selects physical representation independently for each allele row, while H2 represents the paths induced by a specified reference-coordinate catalog. Whether these choices yield advantages over the preceding methods is left to controlled benchmarking. We do not claim equivalence to an arbitrary pangenome graph, sequence reconstruction without allelic sequence payloads, or demonstrated support for downstream clinical decisions.

\section{Data model}
\begin{definition}[Allelic catalog and incidence]
Let $C$ be an ordered catalog of $S$ reference-anchored sites on a chromosome. Site $s$ has an interval $I_s=[p_s,e_s]$ in one-based inclusive coordinates and alleles $\mathcal A_s=\{\mathrm{REF},\mathrm{ALT}_1,\ldots,\mathrm{ALT}_{m_s}\}$. Let $H$ denote a fixed set of phased haplotype identifiers. Define
\[
 x_{s,a,h}=\begin{cases}1,&\text{haplotype $h$ carries allele $a$ at $s$,}\\0,&\text{otherwise,}\end{cases}
\]
with explicit unknown state $x_{s,\unknown,h}$ when the call is unresolved. An H1 row is the set $R_{s,a}=\{h:x_{s,a,h}=1\}$ for a particular nonreference allele $a$. The catalog stores coordinate, allele identity, and any sequence/structural metadata independently of these carrier sets.
\end{definition}
A valid single-site phased call assigns exactly one state from $\mathcal A_s\cup\{\unknown\}$ to each haplotype. An unphased heterozygote specifies a diploid genotype but does not identify its alleles' assignment to the two labeled haplotypes. A conversion that assigns both alleles to both haplotypes is therefore incorrect. In our revised VCF ingestion, unphased heterozygotes are left unresolved; an unphased homozygote may be placed on both haplotypes without ambiguity.

A sample-name intersection between two phased VCF callsets does not establish phase-label consistency. Separate SNV and structural-variant (SV) callsets can differ in phase blocks or in their assignment of the two chromosome copies. Cross-callset haplotype context therefore requires an explicit phase-harmonization check. Allele-level carrier statistics can still be computed separately without asserting a shared long-range phase orientation.

\section{Adaptive H1 encoding}
For a row containing $k$ carriers among $H$ haplotypes, a raw fixed-width sorted-list model requires $k\lceil\log_2 H\rceil$ bits, whereas an uncompressed dense bitmap occupies $H$ bits, before headers and alignment. The idealized crossover is
\begin{equation}
 k^*\approx\frac{H}{\lceil\log_2 H\rceil}.\label{eq:crossover}
\end{equation}
This is a crossover between \emph{two selected code families}; it is not a proof of information-theoretic optimality. For a fixed cardinality $k$ and equally likely carrier subsets, any uniquely decodable lossless description of the carrier set has at least
\begin{equation}
 L_{\min}(H,k)=\log_2\binom{H}{k}\label{eq:lower}
\end{equation}
bits in the counting sense. In particular, a complement list may be more compact than an explicit list for $k$ near $H$. Enumerator codes, Elias--Fano-like representations, run-length methods, or Roaring bitmaps must also be considered \cite{roaring2016}. Equation~\eqref{eq:lower} is not an unconditional lower bound on a structured multirow genomic file that can share redundancy across sites.

A serialized H1 file additionally requires allele identifiers, a mapping between rows and catalog entries, type tags, row offsets, cardinalities (if materialized), and reference/dictionary information. Consequently, we define \emph{physical index size} as the number of bytes occupied by the complete queryable artifact, including its metadata. We distinguish this from (i) estimated row-payload bits, and (ii) a second-stage whole-file Zstandard archive, which may require decompression before queries. These are distinct performance objectives.

\paragraph{Implementation-level selection rule.}
The analytical crossover in Eq.~\eqref{eq:crossover} assumes packed haplotype IDs of width $\lceil\log_2 H\rceil$ bits. The pilot's physical adaptive writer instead uses $32$-bit carrier IDs and selects a list when $4k<\lceil H/8\rceil$ bytes, otherwise a bitset. For $H=400$, this corresponds to $k<12.5$, not the analytical $k^*\simeq44.4$. A separate custom Roaring-like experiment partitions haplotype IDs into blocks and stores array or bitset containers. These implementations are compared as concrete row formats, not as identical realizations of the theoretical bit-cost model. Neither file contains the complete external genomic allele catalog, nucleotide sequences, or the mapping from site identifiers to numerical row IDs; consequently, reported sizes describe incidence-row artifacts rather than full, self-contained genotype databases.

\section{H2 construction and the limits of equivalence}
For fixed chromosome boundaries, let $B=\{b_0<\cdots<b_t\}$ contain the region start, the region end plus one, and each catalog interval's $p_s$ and $e_s+1$. Construct a directed reference edge from boundary $b_i$ to $b_{i+1}$ for each adjacent pair. For each nonreference allele at $s$, add a distinct labeled intermediate vertex $v_{s,a}$ and two edges
\[
 b(p_s)\longrightarrow v_{s,a}\longrightarrow b(e_s+1).
\]
An unknown call can be represented by a distinct unknown branch. The resulting object is an \emph{abstract boundary-edge graph}; nucleotide sequences, orientation and inserted bases do not follow merely from its edge IDs. For any fully called haplotype, an H2 path replaces each selected site's reference subwalk with its chosen two-edge detour, leaving all other reference intervals intact.

\begin{theorem}[Conditional reconstruction of canonical H2 walks]\label{thm:canonical}
Suppose (i) a fixed, ordered reference-coordinate catalog $C$ and its graph-construction rule are known; (ii) every interval $I_s$ lies within the region and distinct sites are pairwise disjoint; (iii) every haplotype has exactly one explicit allele state at every site, with a consistent phase-label convention; and (iv) every required alternative is represented as one atomic catalog allele, with no unmodeled nested or rearranged traversal. Then $C$ and the complete H1 incidence assignment determine one canonical, contiguous H2 boundary-edge walk per haplotype. The walk recovers the original per-site allele assignments by inspection of its alternative branches.
\end{theorem}
\begin{proof}
Order sites by their reference starts. Disjointness implies that the end boundary of site $s$ is no later than the start boundary of the next site. Starting at the region's first boundary, traverse the unique reference edges until the first site. The complete incidence assignment uniquely selects the reference subwalk or one labeled alternative branch at that site. Both choices reach its end boundary, from which reference edges lead to the next site. Induction over the ordered sites yields a contiguous directed walk, with no choice left unspecified by the assignment. After the final site, reference edges uniquely connect to the region's end. The two-edge alternative branch uniquely identifies each nonreference allele, whereas reference-edge traversal identifies the reference allele. Thus, the per-site assignment can be recovered. This establishes a bijection between valid assignments and walks in the \emph{canonical catalog-induced construction}, not between incidence and arbitrary genomic graphs.
\end{proof}

\begin{proposition}[Failure of unrestricted graph equivalence]
Allele--haplotype incidence without a fixed catalog or with arbitrary overlapping/rearranged variation does not uniquely determine the node-labeled topology or sequence content of a pangenome graph.
\end{proposition}
\begin{proof}
Given the same single-site incidence matrix, one may construct graphs that split a reference segment into different numbers of vertices, contract a linear chain, or attach distinct sequence labels, without changing which haplotypes carry the catalog allele. These graphs differ in their labeled topology while yielding identical incidence. Overlapping variants introduce an additional ambiguity: selecting two alternatives need not define a coherent reference-coordinate traversal. Therefore incidence alone cannot select a unique graph or guarantee a valid sequence realization.
\end{proof}

\begin{remark}[SV semantics]
Deletions, insertions, inversions, duplications, and breakends are not interchangeable sequence operations. The present H2 edge graph uses atomic allelic alternatives to describe carrier choice. It does not, without a richer catalog and orientation/sequence semantics, establish nucleotide-level reconstruction of inversions, tandem copy number, or translocations. Breakend/translocation records are currently excluded from the exact-path construction.
\end{remark}

\section{Supported operations and predicted costs}
Let $w$ be the machine word width. A sparse sorted-list H1 row of cardinality $k$ supports sequential carrier enumeration in $O(k)$ time; a dense raw bitmap requires $O(\lceil H/w\rceil)$ word inspection in the naive implementation, although optimized bit iteration can reduce output overhead. Sparse--sparse intersection by a merge requires $O(k_1+k_2)$ time. Dense--dense intersection needs $O(\lceil H/w\rceil)$ word operations. A physical compressed container layout changes these constants and may change the optimal algorithm. Carrier frequency follows from stored cardinalities or a bitmap popcount. An H2 path gives explicitly ordered edges, but path reconstruction, random access, and adjacency queries must be timed on the encoded representation; building an in-memory inverted index duplicates some information and is not automatically storage-free.

\section{Experimental methods and results}

\subsection{Inputs, sampling, and variant representation}
The completed local notebook uses two phased high-coverage 1000 Genomes Project callsets distributed through the International Genome Sample Resource~\cite{byrska2022}. The SNV/indel input is the NYGC 30$\times$ phased callset, and the SV-containing input is the 1kGP high-coverage Illumina phased panel. Their exact filenames are recorded in the accompanying result-provenance manifest. Records are restricted to chromosome 1, positions 20,000,000--22,000,000 (one-based inclusive), and a reproducibly shuffled subset of 200 sample identifiers selected using Python's \texttt{random.Random(42)}. The subset is reused in both readers. The SV layer retains records bearing the \texttt{SVTYPE} annotation; no joint phase-harmonization method is applied to the two input callsets.

The saved run reports 24,921 SNV/indel records, 45 SVTYPE-annotated records, 400 labeled haplotypes, and 24,966 merged sites (Table~\ref{tab:catalog}). The all-alternative-row extraction produces 24,966 nonreference incidence rows for this dataset. In this experiment the number of extracted rows therefore equals the number of retained sites, but that equality is not a property of multiallelic catalogs generally. Allele labels, genomic positions, and sequence or structural annotations remain external to the physical incidence-row files.

\subsection{Restricted H2 projection and correctness checks}
The canonical H2 construction selects a nonoverlapping subset of the SNV/indel catalog. The saved notebook reports 22,727 retained sites and 2,194 excluded overlaps (8.8\% of the original SNV/indel input). A separate SV-only demonstration retains 40 of 45 records and excludes five. The two projections are not concatenated into a single verified chromosome-wide H2 traversal. The disjoint-site procedure imposes a representation restriction rather than resolving all alternative SV semantics.

Within the retained SNV projection, the notebook reports that every 22,727 mapped alternative row passes H1--H2 carrier-set assertions and that consecutive edges satisfy the abstract graph-adjacency invariant. Demonstration queries recover equal carrier sets for an allele (97 in both views), equal intersections for a selected pair (zero in both views), and identical membership for one selected haplotype. Additional tests establish exact equality of examined raw and decoded Huffman path prefixes and selected path-record round-trips. These are internal consistency checks, not independent validation of nucleotide sequences or all external phase relationships. The earlier, pre-correction microsecond timings are not used as comparative performance evidence.

\subsection{Physical storage experiment}
The H1 experiment writes carrier rows with two locally implemented layouts: a simple per-row choice between a 32-bit integer list and a $\lceil H/8\rceil$-byte bitmap, and a custom Roaring-like block/container file. Both include their own headers, payload structures, and row offsets, but omit the mapping from numeric row IDs to allele names and other genomic catalog metadata. The notebook provides a disk reader for the custom Roaring-like format; it does not yet demonstrate a complete read/query/round-trip interface for the simpler adaptive file. The reference bitmap baseline is a payload calculation, not a complete disk index. The optional Zstandard pass compresses the adaptive row file as one archive; its size does not imply direct random-access queries.

For H2, the notebook writes indexed sequences of integer graph-edge IDs from the same SNV-only canonical projection, once as 32-bit records and once using a context-dependent canonical Huffman model with indexed access. The edge-map and full sequence catalog are not included in the measured files. File size was obtained from local \texttt{os.path.getsize} calls, not estimated from theoretical entropy. Comparisons among these restricted files are meaningful; comparisons with BCF, GBWT, GQT, BGT, GRG or other full genotype systems have not been executed.

\subsection{Recorded storage results}
\begin{figure}[htbp]
\centering
\includegraphics[width=0.92\linewidth]{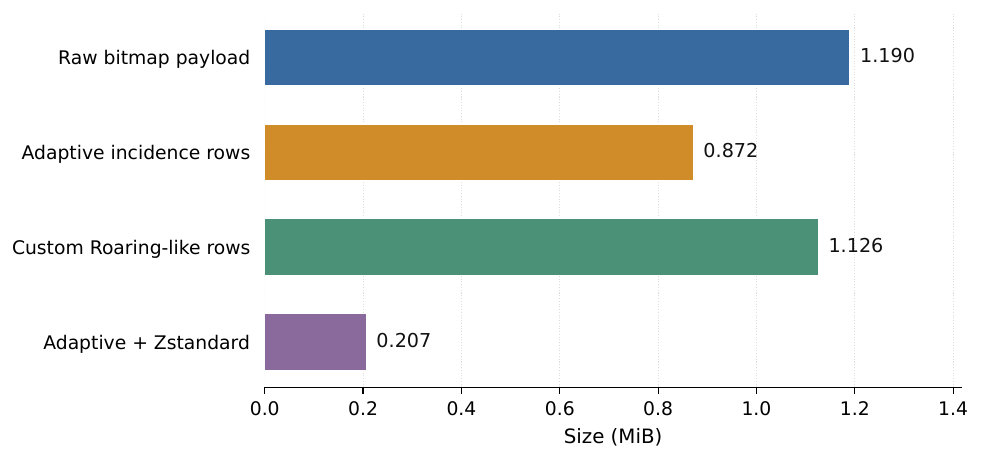}
\caption{Recorded chromosome 1 H1 incidence-row storage. The uncompressed bitmap baseline excludes row offsets and genomic metadata. The Zstandard archive is distinct from a queryable index.}
\label{fig:h1size}
\end{figure}

Table~\ref{tab:h1files} and Fig.~\ref{fig:h1size} summarize the local notebook's H1 measurements. Among the 24,966 rows, 16,963 (67.9\%) select the sparse representation and 8,003 (32.1\%) select a dense bitmap under the physical writer's $4k$-byte list rule. The 0.872 MiB adaptive file is 1.37 times smaller than the bitmap-payload baseline, and the custom Roaring-like file is 1.06 times smaller. The adaptive file archived with Zstandard occupies 0.207 MiB, but is not a queryable row index in the tested notebook. The two row-file formats differ from a production Roaring implementation, and their byte counts should not be interpreted as a benchmark against one. The smaller adaptive file was measured by the notebook writer, but a complete serialized-row decoding audit remains outstanding.

\begin{figure}[htbp]
\centering
\includegraphics[width=0.92\linewidth]{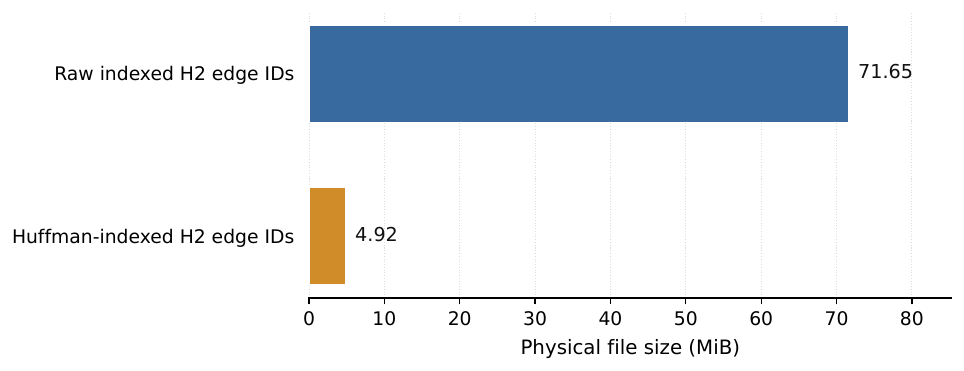}
\caption{Recorded indexed H2 edge-ID sequence storage for the 22,727-site nonoverlapping SNV projection. The measured files omit the external edge-to-allele catalog and nucleotide payload.}
\label{fig:h2size}
\end{figure}

Table~\ref{tab:h2files} and Fig.~\ref{fig:h2size} report the analogous local indexed edge-stream comparison for H2. The raw file occupies 75,130,882 bytes and the context-Huffman version 5,156,418 bytes, giving a ratio of 14.57. This is an edge-stream encoding result conditional on the chosen canonical graph; it is not 14.57-fold compression of genomic sequences, or 14.57-fold improvement over an established haplotype index.

\subsection{Synthetic scaling and unperformed comparative experiments}
The notebook also simulates larger $H$ by holding the observed per-row carrier counts $k$ fixed and resampling carrier IDs uniformly without replacement. This synthetic design preserves the \emph{absolute} carrier-count spectrum but not observed population frequencies, linkage disequilibrium, inheritance, or the number of newly observed alleles. For example, its reported crossover in format ordering near $H=3{,}200$ and its high-$H$ compression ratios are properties of that fixed-$k$ construction, not biological cohort-scaling evidence. We therefore exclude these synthetic extrapolations from the primary results.

Four outstanding checks define the next publication stage: (i) end-to-end artifact verification against genotype and allele metadata, including all decoded rows; (ii) normalization of overlapping variants and cross-callset phase analysis; (iii) comparisons against identical-payload BCF and production queryable genotype/pangenome indexes; and (iv) real-cohort scaling with build time, peak memory, median/tail query latency, and independent execution-environment records. These have not been completed and are not inferred from the present internal pilot.

\begin{table}[htbp]
\centering
\caption{Local chromosome 1 pilot and restricted H2 projections. The SNV-only and SV-only projections are separate; neither establishes phase alignment across the merged catalogs.}
\label{tab:catalog}
\begin{tabular}{@{}lrrr@{}}
\toprule
Input / projection & Sites retained & Excluded & Haplotypes\\
\midrule
SNV/indel H1 input & 24,921 & --- & 400\\
SVTYPE-annotated H1 input & 45 & --- & 400\\
Merged H1 site catalog & 24,966 & --- & 400\\
Canonical H2 from SNV/indel input & 22,727 & 2,194 & 400\\
Canonical H2 from SV-only input & 40 & 5 & 400\\
\bottomrule
\end{tabular}
\end{table}

\begin{table}[htbp]
\centering
\caption{Recorded physical H1 incidence-row sizes from the 400-haplotype pilot. MiB = $2^{20}$ bytes. The baseline comprises only $24{,}966\lceil400/8\rceil$ bitmap bytes. All formats omit the separately stored genomic catalog. The final row is a whole-file archive and is not directly queryable by the notebook's row-index interface.}
\label{tab:h1files}
\begin{tabular}{@{}lrr@{}}
\toprule
H1 representation & Size (MiB) & Baseline / size\\
\midrule
Raw bitmap payload (baseline) & 1.190 & 1.00\\
Adaptive row file (\texttt{h1\_adapt.raw}) & 0.872 & 1.37\\
Custom Roaring-like file (\texttt{h1\_roar.rd}) & 1.126 & 1.06\\
Adaptive row file + Zstandard & 0.207 & 5.75\\
\bottomrule
\end{tabular}
\end{table}

\begin{table}[htbp]
\centering
\caption{Recorded indexed H2 edge-stream storage, using only the 22,727-site nonoverlapping SNV projection. These files do not store the underlying nucleotide sequence catalog or a self-contained graph topology description.}
\label{tab:h2files}
\begin{tabular}{@{}lrr@{}}
\toprule
H2 indexed edge-stream format & Bytes & MiB\\
\midrule
Raw 32-bit edge IDs & 75,130,882 & 71.65\\
Context-Huffman coded edge IDs & 5,156,418 & 4.92\\
\bottomrule
\end{tabular}
\end{table}

\section{Threats to validity and interpretation}
The principal structural limitation is the nonoverlap requirement: the canonical SNV H2 projection excludes 2,194 of 24,921 sites, and the separate SV demonstration excludes five of 45 records. Thus, even exact carrier equivalence within the projected graph does not imply preservation of the complete merged catalog. Cross-callset phase consistency has not been demonstrated, and symbolic SV records lack the sequence and orientation semantics required for nucleotide-level reconstruction. Internal equality assertions do not constitute biological validation of the underlying calls.

The H1 bitmap comparator is payload-only while the adaptive and custom Roaring-like figures include row metadata, and none of those files includes the external genomic allele catalog or nucleotide sequence information. The simple adaptive file has no independently exercised on-disk query/reader benchmark in the saved notebook. The compressed H2 format encodes edge-ID streams rather than the full graph representation. Consequently, neither an H1 size ratio nor an H2 raw-to-Huffman ratio should be interpreted as a direct comparison with complete BCF, GBWT, GQT, BGT, GRG, or other production systems. The synthetic cohort expansion holds carrier cardinalities fixed and cannot substitute for replication using additional real samples. The old microbenchmark timings are invalidated by the changed H2 construction and are not reported as speed claims.

A positive or negative operational comparison would remain specific to the tested payloads, variant classes, and query workloads. No claim is made concerning improved genome alignment, clinical interpretation, pharmacogenomics, drug discovery, privacy, or lossless reconstruction of arbitrary pangenome graph topology.

\section{Conclusion}
We presented an adaptive allele-incidence row format (H1) and a conditionally reconstructable abstract edge-walk representation (H2). The local 400-haplotype pilot records a 0.872 MiB adaptive incidence-row file from 24,966 allele rows and 14.57-fold compression of indexed H2 edge streams relative to raw edge-ID storage on a separate 22,727-site nonoverlapping SNV projection. The row files omit genomic catalogs; the H2 stream omits graph sequence payloads; and the two measurements cannot be combined into a complete compressed pangenome benchmark. Internal carrier and path-adjacency assertions support the restricted construction, but neither full SV reconstruction nor superiority over established indexes follows. The next decisive result must compare equivalent end-to-end payloads and query workloads using reproducibly validated real data.

\section*{Code, data, and reproducibility status}

The experimental evidence presented in this study is derived from
the saved outputs of the completed local H1/H2 analysis notebook
(28 October 2026). The notebook documents the input dataset
identifiers, genomic region, deterministic 200-sample selection,
construction procedures, internal consistency checks, physical
storage measurements, and figure-generation workflow.

The associated research software and reproducibility materials
have been deposited in a public, versioned Zenodo archive,
available at \url{https://doi.org/10.5281/zenodo.23727279}.
The archive supports inspection of the computational methodology,
experimental implementation, and reported results.

The original genomic VCF datasets are not redistributed with
the manuscript; their availability remains governed by the
respective data providers. The reported H1 storage measurements
exclude the external allele catalog, whereas the H2 measurements
concern a restricted nonoverlapping SNV projection and exclude
external graph metadata and nucleotide payloads. Synthetic
cohort-scaling experiments are distinguished from measurements
on genomic data.

Although the public release improves transparency and
reproducibility, independent replication of the reported
numerical measurements and comprehensive comparisons with
established production genomic indexes have not yet been
completed. Consequently, the reported compression results
should be interpreted as experimental measurements within
the stated implementation and evaluation conditions, rather
than evidence of general superiority over existing methods.

\end{document}